\documentstyle[12pt]{article}

\input{epsf.sty}
\def\framegraphics{\def\ifframe{\iftrue}}
\def\dontframegraphics{\def\ifframe{\iffalse}}
\def\drawgraphics{\def\ifdraw{\iftrue}}
\def\dontdrawgraphics{\def\ifdraw{\iffalse}}

\dontframegraphics
\drawgraphics

\newcommand{\graphics}[6]{
\def\epsfsize##1##2{#6##1}
\begin{picture}(#2,#3)
  \ifframe
    \put(0,0){\framebox(#2,#3){}}
  \fi
  \ifdraw
    \put(0,#3){\begin{picture}(0,0)
                 \put(#4,#5){\epsfbox{#1}}
               \end{picture}}
  \fi
\end{picture}}

\newcommand{\be}{\begin{equation}}
\newcommand{\ee}{\end{equation}}
\newcommand{\bear}{\begin{eqnarray}}
\newcommand{\ear}{\end{eqnarray}}

\begin{document}


\vspace{2cm}
\centerline{A Comparative Study for the Evolution of Structure 
Functions: DGLAP versus FOPT}
\vspace{.4cm}
\centerline{F.M. Steffens\footnote{Email address:fsteffen@axpfep1.if.usp.br}}
\vspace{1cm}
\centerline{Instituto de F\'{\i}sica}
\centerline{Universidade de S\~ao Paulo}
\centerline{C.P. 66318}
\centerline{05389-970, S\~ao Paulo, Brazil}
\vspace{.3cm}
\begin{abstract}
We investigate an alternative to the DGLAP evolution of 
structure functions through the use of Fixed Order Perturbation
Theory. Remarkable agreement between the two methods are 
found in the polarized sector for a wide $x$, $Q^2$ region.
However, for the unpolarized sector the agreement is poorer.
\end{abstract}

In the last few years we have seen an explosion of new and 
precise data for unpolarized \cite{unpo} and polarized \cite{pol} 
structure functions. These experimental data cover a wide kinematical 
range, with $10^{-5}<x<0.85$
and $0.2<Q^2<5000 \; GeV^2$ for the unpolarized case and with $0.003<x<0.7$
and $1.3 < Q^2<58 \; GeV^2$ for the polarized case. Along with the tremendous 
experimental advances, we have also seen firm theoretical developments in 
the sector of perturbative QCD \cite{levin,forte}.
For instance, nowadays we have at our disposition the complete set of the
polarized next-to-leading order (NLO) anomalous dimensions 
\cite{mertig96,zijlstra94}. 
These developments, together with the accurate
data, have allowed us to perform a number of analysis with NLO precision 
of the behaviour of the parton distributions and of the structure functions
\cite{param1,param2}. 
However, for practical applications outside the published 
parametrizations for the data one needs to write a somewhat long and 
involved program to solve the DGLAP \cite{dglap} equations for QCD
evolution of the parton distributions. 
In doing so one meets problems of all sorts, from
lengthy expressions for the anomalous dimensions in NLO to worries about
analytical continuation, etc. These sorts of technical difficulties
make it hard for those who want a straightforward and more direct way
to study evolution and are not directly
involved with these programs. Moreover, the extension from the 
present day NLO calculations to the full computation 
of the NNLO anomalous dimensions will be, per se, a giant task.
Hence, a more simple approach should 
be available, at least for the evolution of the structure
functions. In this work, we explore an alternative
to the evolution of structure functions via a systematic study
of what is known as  Fixed Order Perturbation 
Theory (FOPT) \cite{zijlstra94}. 
There is also a related interest in FOPT when dealing with 
heavy quark production \cite{aivazis,buza,steffens,roberts}. Although we
do not treat this problem here, our considerations are relevant
because they address exactly the main point of heavy quark production:
when to remove the mass-singular logarithms of the coefficient 
functions? 
Following this point, we make a brief  
discussion of the problem of logarithm resummation and its necessity. 
This is a problem very much in vogue today because of the
apparent lack of necessity of resummation of the $1/x$ logarithms,
as seen from the HERA data. In our specific case, we will see that these
$1/x$ singularities, apparently disconnected from what we are investigating
here (the scale logarithms), may play a fundamental role. 

Structure functions are cross sections which can be calculated with the 
help of the factorization theorem:

\be
F(x, Q^2) = \sum_{f=q,g} \int_x^1 \frac{dz}{z} f\left(\frac{x}{z},\mu^2\right)
\left[\delta (z-1)\delta_{fq}
+ \frac{\alpha_s (\mu^2)}{4\pi} C_f^{(1)} \left(z, \frac{Q^2}{\mu^2}\right) 
+ ...\right],
\label{a1}
\ee
where $F(x, Q^2)$ is a general structure function and expression 
(\ref{a1}) is saying
that the full cross section can be separated into a perturbative part,
the term in brackets, and a nonperturbative part, the parton distribution 
$f (x/z)$. The renormalization and mass factorization scales 
are chosen to be the same and equal to 
$\mu$. The delta function $\delta_{fq}$ reflects the fact that 
the gluon contribution starts only at order $\alpha_s$.

Because of the infrared singularity, the coefficients $C^{(1)}$ depend 
on $ln(Q^2/\mu^2)$. Once we fix $\mu^2$ we can
calculate $F (x, Q^2)$ at any $Q^2$. In this sense, the evolution
of $F (x, Q^2)$ is made through the logarithms appearing in $C^{(1)}$.
However, this way to evolve the structure function is usually considered 
to be inadequate because the logarithms diverge when $Q^2 >> \mu^2$. 
To avoid this divergence, the logarithms are then resummed through the 
renormalization group equations and $F (x, Q^2)$ is given by:

\be
F(x, Q^2) = \sum_{f=q,g} \int_x^1 \frac{dz}{z} f\left(\frac{x}{z}, Q^2\right)
\left[\delta (z-1)\delta_{fq}
+ \frac{\alpha_s (Q^2)}{4\pi} C_f^{(1)} (z) + ...\right].
\label{a2}
\ee
The calculation of $F(x, Q^2)$ through Eq. (\ref{a1}) is what is meant 
by FOPT evolution in this paper, and it is quite simple to implement
this program; a few lines in a Mathematica program will do it.
One only needs to calculate the relevant photon-parton
cross sections to the required order. On the other hand,
the computation of $F(x, Q^2)$ through Eq. (\ref{a2}) requires
the knowledgement not only of the coefficient functions $C_f^{(1)} (z)$
but also the singlet and non-singlet anomalous dimensions. Besides that,
one needs to solve the corresponding Altarelli-Parisi evolution equations for 
the parton distributions $f(x/z)$, which may turn out to be a non trivial
task \cite{gluck90,wally}. 
The question is then: which are the limits of applicability,
or of equivalence, of equations (\ref{a1}) and (\ref{a2})?

To answer this question, we start considering the $F_{2p}$ structure
function for four flavours in FOPT:

\bear
F_{2p} (x, Q^2) &=& \frac{5}{18} \int_x^1 \frac{dz}{z} \left[ q^{S}\left(
\frac{x}{z}, \mu^2\right)
C_q^{S} (z, Q^2/\mu^2) + g(x/z, \mu^2) 
C_g (z, Q^2/\mu^2)\right] \nonumber \\*
&+&\frac{1}{6} \int_x^1 \frac{dz}{z} q^{NS} (x/z, \mu^2)
C_q^{NS}(z, Q^2/\mu^2).
\label{a3}
\ear
The singlet distribution is $q^S(z) = u(z) + \overline u(z)
+ d(z) + \overline d(z) + s(z) + \overline s(z) + c(z) + \overline c(z)$,
and the nonsinglet distribution is $q^{NS}(z) =  u(z) + \overline u(z) -
 d(z) - \overline d(z) - s(z) -\overline s(z) - \overline d(z) + c(z) 
+ \overline c(z)$.
As we are studying NLO evolution of parton distributions, the 
coefficient functions are necessary to be known only to order $\alpha_s$.  
The non-singlet quark coefficient is \cite{zijlstra92}

\bear
C_q^{NS} (z, Q^2/\mu^2) &=& \delta (z-1) + \frac{\alpha_s(\mu^2)}{4\pi}\left\{
C_F\left(\frac{4}{1-z} ln(1-z) - \frac{3}{1-z}
+ \frac{4}{1-z} ln\left(\frac{Q^2}{\mu^2}\right)\right)_+ \right. \nonumber \\* 
&-& \left. C_F\left[2(1+z)ln(1-z) - 2\frac{1+z^2}{1-z}ln z + 6 + 4z 
- 2(1 + z)ln\left(\frac{Q^2}{\mu^2}\right)\right] \right. \nonumber \\*
&+& \left. C_F\left[-4\zeta(2) - 9 + 3 ln\left(\frac{Q^2}{\mu^2}\right)
\right]\delta (z-1)\right\},
\label{a4}
\ear 
and to this order, $C_q^S(z, Q^2/\mu^2) = C_q^{NS} (z, Q^2/\mu^2)$. 
The gluon coefficient is given by \cite{zijlstra92}:

\bear
C_{g} (z, Q^2/\mu^2) &=& n_f T_f \frac{\alpha_s(\mu^2)}{4\pi}\left\{
4 (1 - 2z + 2z^2)ln\left(\frac{Q^2}{\mu^2}\right)  
- 4 + 32z(1-z) \right. \nonumber \\*
&+&\left. 4(1 - 2z + 2z^2)ln\left(\frac{1-z}{z}\right) \right\}.
\label{a5}
\ear
For the quark distributions, we choose the  CTEQ3 \cite{cteq3} 
parametrization, valid at $Q^2 = 2.56\; GeV^2$. It is then very
clear why FOPT is so simple: Eqs. (\ref{a3}) - (\ref{a5}) is all 
we need to calculate $F_{2p} (x, Q^2)$ for any $x$, $Q^2$ pair.

\begin{figure}[htb]
\graphics{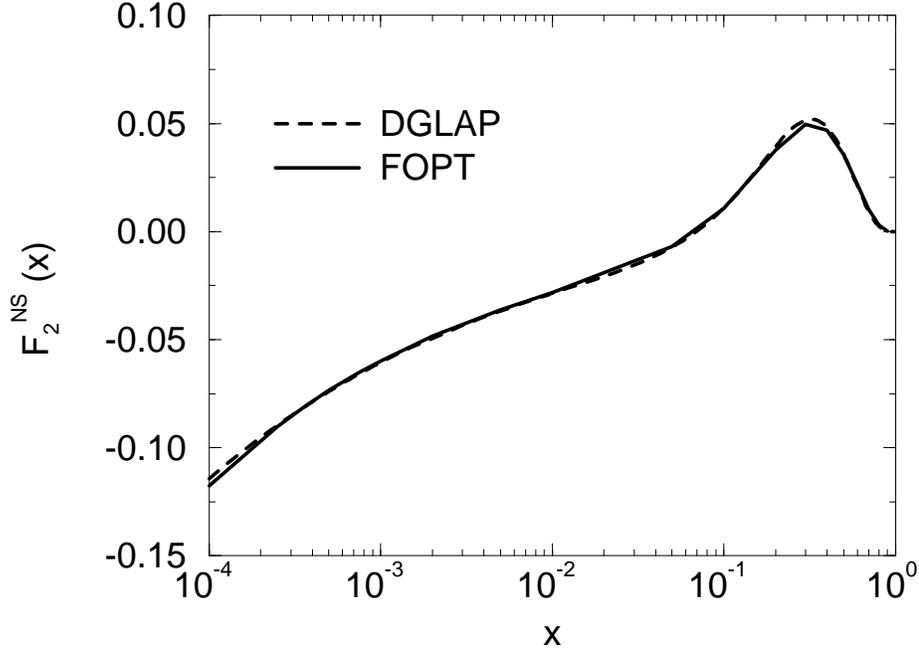}{20}{10}{-2}{-10}{0.9}
\caption{The non-singlet part of $F_{2p}$. The evolution is 
from 2.56 to $10 GeV^2$. The continuous and dashed lines are,
respectively, the FOPT and DGLAP results at $10 GeV^2$.}
\label{fig1}
\end{figure}

\begin{figure}[htb]
\graphics{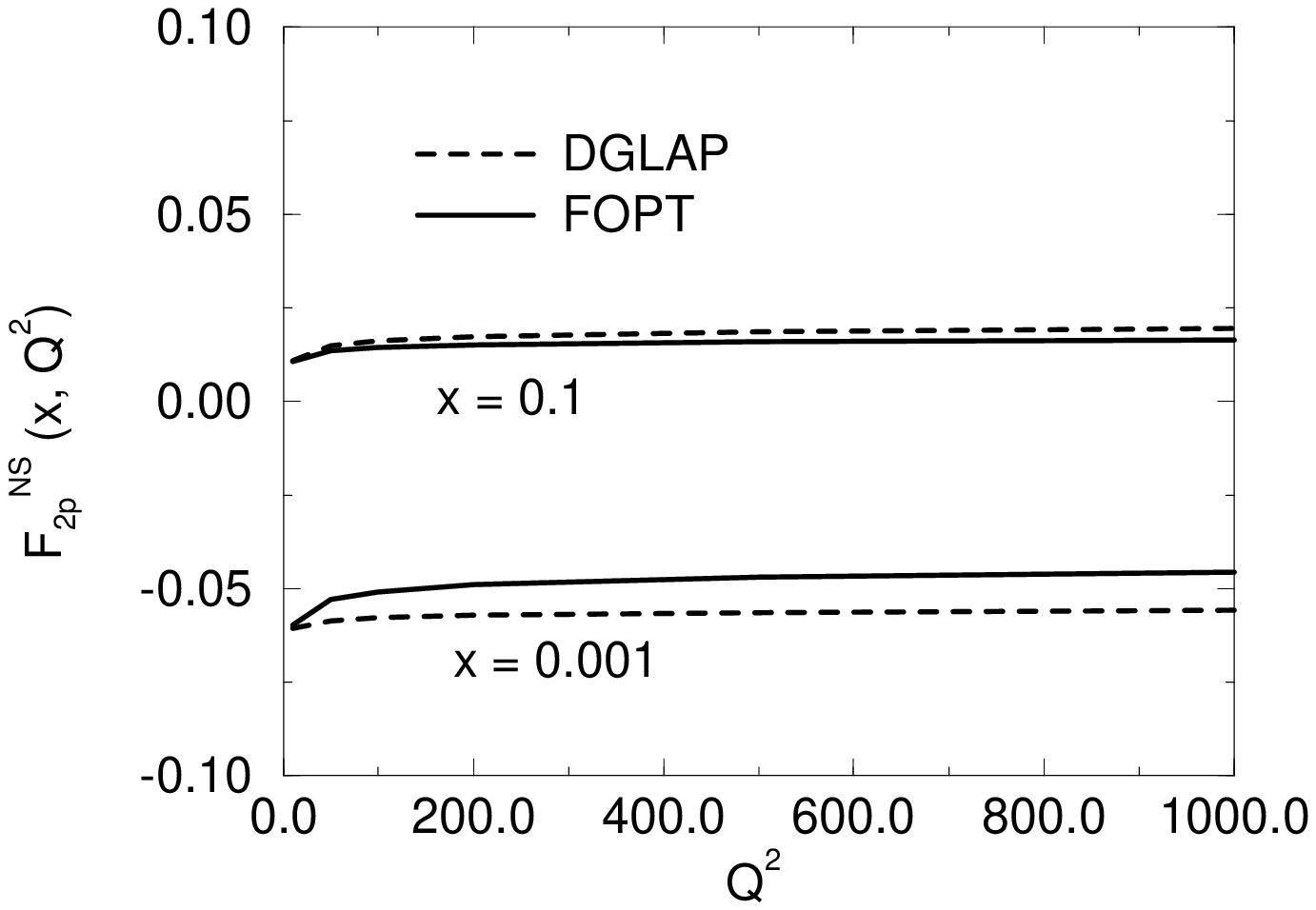}{20}{10}{-2}{-10}{0.9}
\caption{The $Q^2$ dependence of $F_{2p}^{NS}$, for two specific
values of $x$. The initial scale for evolution is 2.56 $GeV^2$.}
\label{fig2}
\end{figure}

We now present the behaviour of $F_{2p} (x, Q^2)$ with $x$ and
$Q^2$ for FOPT and DGLAP evolution. We start showing in 
Figure \ref{fig1} the non-singlet piece, evolved 
from $\mu^2 = 2.56\; GeV^2$ to $Q^2 = 10\; GeV^2$. The continuous line
is the evolution result using FOPT and the dashed line is the 
result using DGLAP. They totally agree down to $x = 10^{-4}$. 
Of course, at $10\; GeV^2$, $ln(Q^2/\mu^2) = 1.36$, which is 
in principle a small number. In this case, there would not be
necessary any resummation of these logarithms, something which
would be relevant only when $Q^2 >> \mu^2$. Thus, the interesting
point here is 
to study the behaviour of $F_{2p}^{NS} (x, Q^2)$ with $Q^2$. That is shown in 
Figure \ref{fig2} for two specific points: $x = 0.1$ and $x=0.001$. 
We see that even for $Q^2 = 1000\; GeV^2$, FOPT and DGLAP still give 
very close results, questioning the necessity of logarithm resummation 
for the nonsinglet piece of $F_{2p}$. 

\begin{figure}[htb]
\graphics{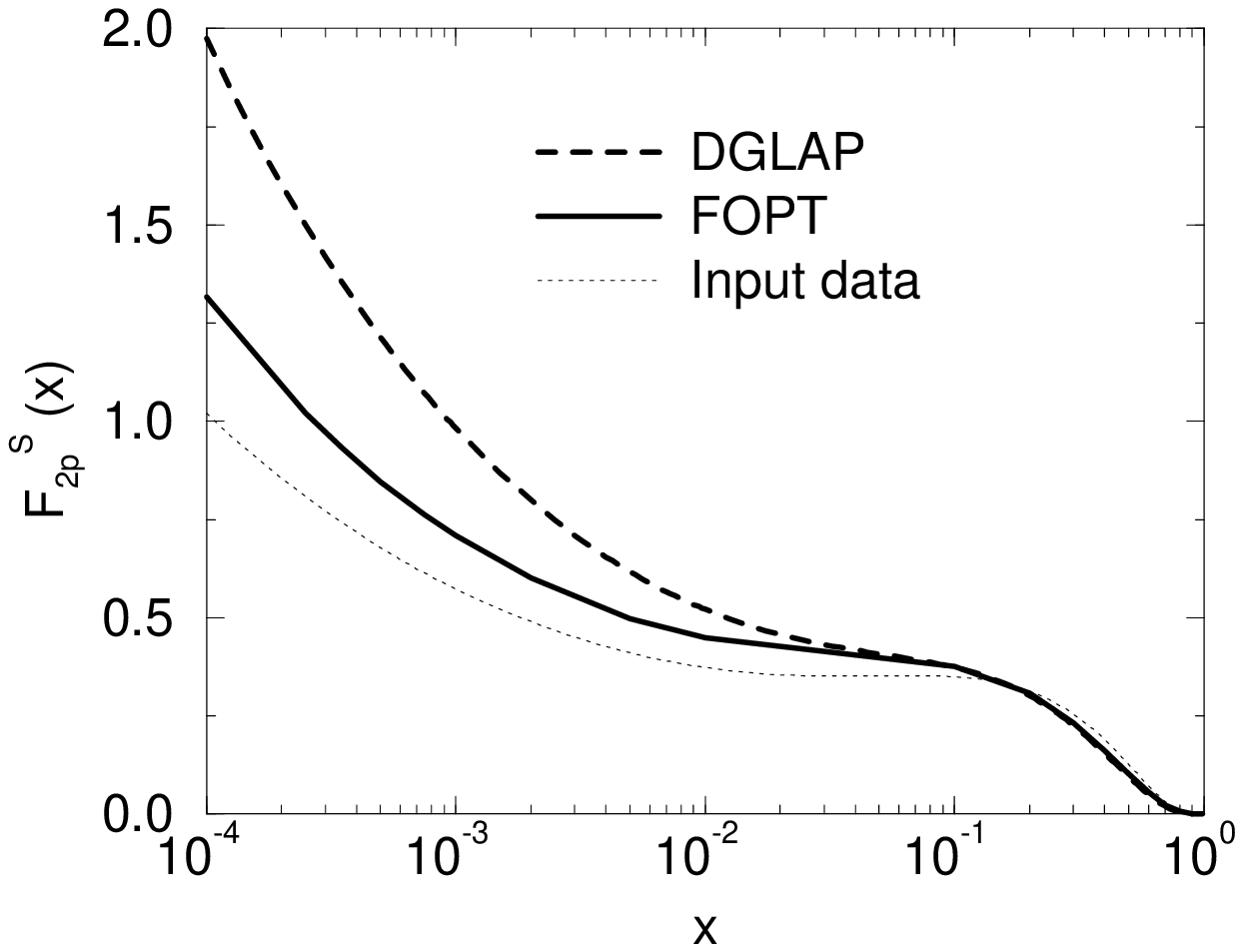}{20}{10}{-2}{-10}{0.9}
\caption{The singlet part of $F_{2p}$. The evolution is 
from 2.56 to 10 $GeV^2$. The continuous and dashed lines are,
respectively, the FOPT and DGLAP results at 10 $GeV^2$. The dotted
line is the input distribution at 2.56 $GeV^2$.}
\label{fig3}
\end{figure}

In the singlet sector the agreement between FOPT and DGLAP is much
poorer. As shown in Figure \ref{fig3}, the two methods 
become inconsistent for $x$ below $0.01$\footnote{In an earlier work
\cite{zijlstra94}, it was concluded that FOPT and DGLAP evolution
were equivalent down to $x=0.01$. However, those authors did not
extended their analysis to lower values of $x$.}. The origin of this
discrepancy should not be in the logarithm involving the scale, 
as it is a small number for the evolution from 2.56 to 10 $GeV^2$.
This result together with considerations made throughout this paper implies, 
contrary to general belief, that the necessity of resummation may 
not be directly connected to the size of the logarithm. At least
in the case of the logarithms involving the scale.
A discussion of this point, along with a possible interpretation
to it, is made after the polarized evolution is studyed.

\begin{figure}[htb]
\graphics{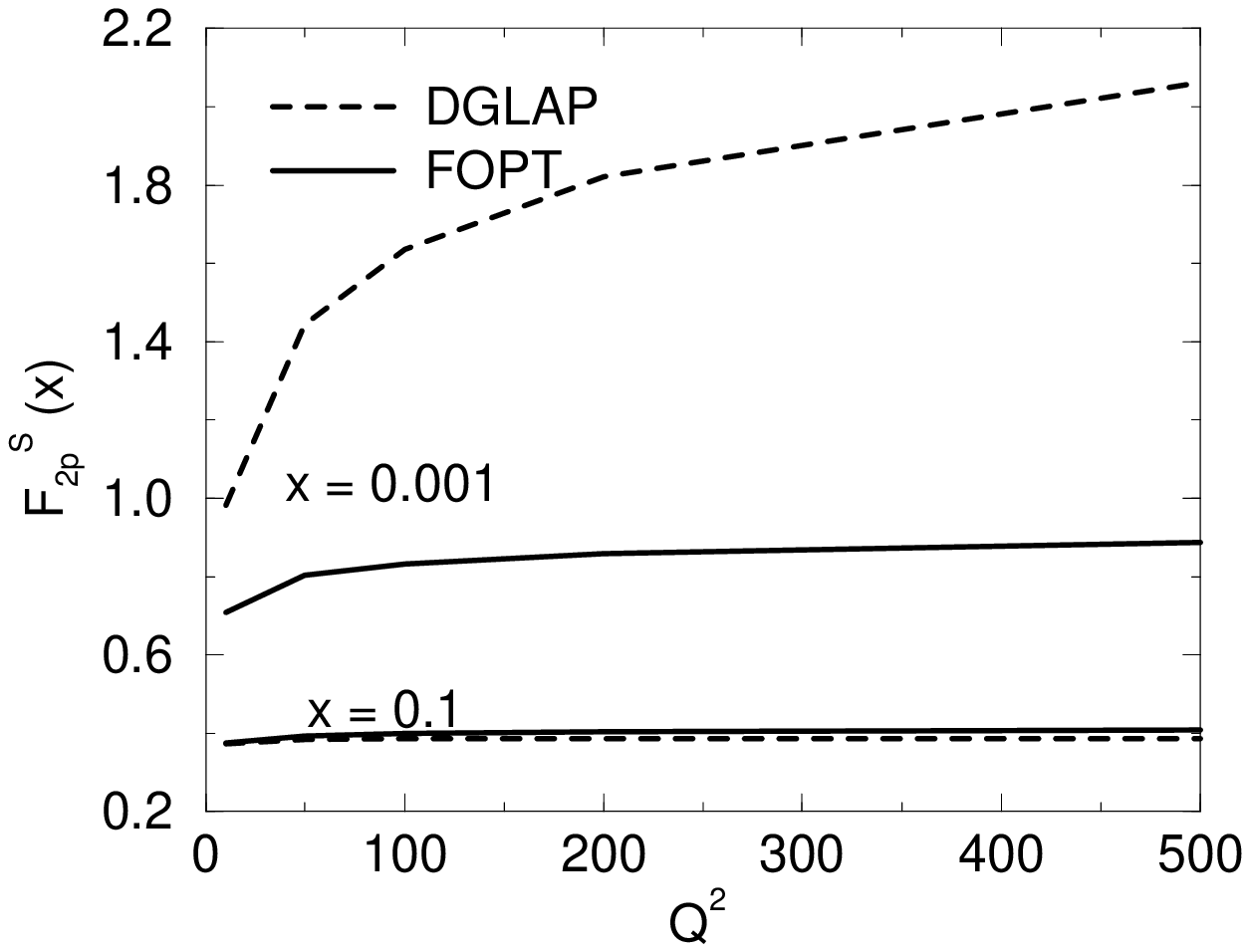}{20}{10}{-2}{-10}{0.9}
\caption{The $Q^2$ dependence of $F_{2p}^{S}$, for two specific
values of $x$. The initial scale for evolution is 2.56 $GeV^2$.}
\label{fig4}
\end{figure}

When varying the scale, there is no significant
change for points where the agreement between DGLAP and FOPT was
already good at 10 $GeV^2$. However, as can be seen in
Figure \ref{fig4}, for points like $x = 10^{-3}$
there is a strong $Q^2$ dependence for the DGLAP evolution while
only a very weak dependence for FOPT. In effect, already at 
100 $GeV^2$ the two methods give results which differ by 
100 $\%$. 

Our next step is to  study the 
compatibility between FOPT and DGLAP evolution in the polarized
sector. The polarized structure function $g_{1p} (x, Q^2)$ is given 
in FOPT by:

\bear
g_{1p} (x, Q^2) &=& \frac{5}{36} \int_x^1 \frac{dz}{z} 
\left[\Delta q^{S} (x/z, \mu^2)\Delta C_q^{S}(z, Q^2/\mu^2) 
+ \Delta g(x/z, \mu^2) \Delta C_g (z, Q^2/\mu^2)\right] 
\nonumber \\*
&+&\frac{1}{12} \int_x^1 \frac{dz}{z} \Delta q^{NS} (x/z, \mu^2)
\Delta C_q^{NS}(z, Q^2/\mu^2).
\label{a6}
\ear
The photon-parton cross sections have to be calculated to order $\alpha_s$, as
we are again considering only $\alpha_s$ corrections to the structure 
functions. Hence, the non-singlet quark coefficient is \cite{zijlstra94}:

\bear
\Delta C_q^{NS} (z, Q^2/\mu^2) &=& \delta (z-1) + \frac{\alpha_s(\mu^2)}{4\pi}\left\{
C_F\left(\frac{4}{1-z} ln(1-z) - \frac{3}{1-z}
+ \frac{4}{1-z} ln\left(\frac{Q^2}{\mu^2}\right)\right)_+ \right. \nonumber \\* 
&-& \left. C_F\left[2(1+z)ln(1-z) - 2\frac{1+z^2}{1-z}ln z + 4 + 2z 
- 2(1 + z)ln\left(\frac{Q^2}{\mu^2}\right)\right] \right. \nonumber \\*
&+& \left. C_F\left[-4\zeta(2) - 9 + 3 ln\left(\frac{Q^2}{\mu^2}\right)
\right]\delta (z-1)\right\}.
\label{a7}
\ear 
To the order considered, $C_q^S (z, Q^2/\mu^2) = C_q^{NS} (z, Q^2/\mu^2)$ 
and the gluon coefficient is given by \cite{zijlstra94}:

\bear
\Delta C_{g} (z, Q^2/\mu^2) &=& n_f T_f \frac{\alpha_s(\mu^2)}{4\pi}\left\{
4(2z - 1)ln\left(\frac{Q^2}{\mu^2}\right)  
+ 4(31-4z) \right. \nonumber \\*
&+&\left. 4(2z - 1)ln\left(\frac{1-z}{z}\right) \right\}.
\label{a8}
\ear

We use as input data the same parametrizations used
in the unpolarized case, i.e.,  
$\Delta q^{NS} (x) = q^{NS} (x)$, $\Delta q^{S} (x) = q^S (x)$
and $\Delta g (x) = g(x)$. The only difference in the evolution
of the polarized and of the unpolarized structure functions is
in the coefficient functions. We are completely free to do this 
choice here as we are not aiming at a description of the experimental
data but only trying to determine the range of compatibility 
between FOPT and DGLAP evolution.

\begin{figure}[htb]
\graphics{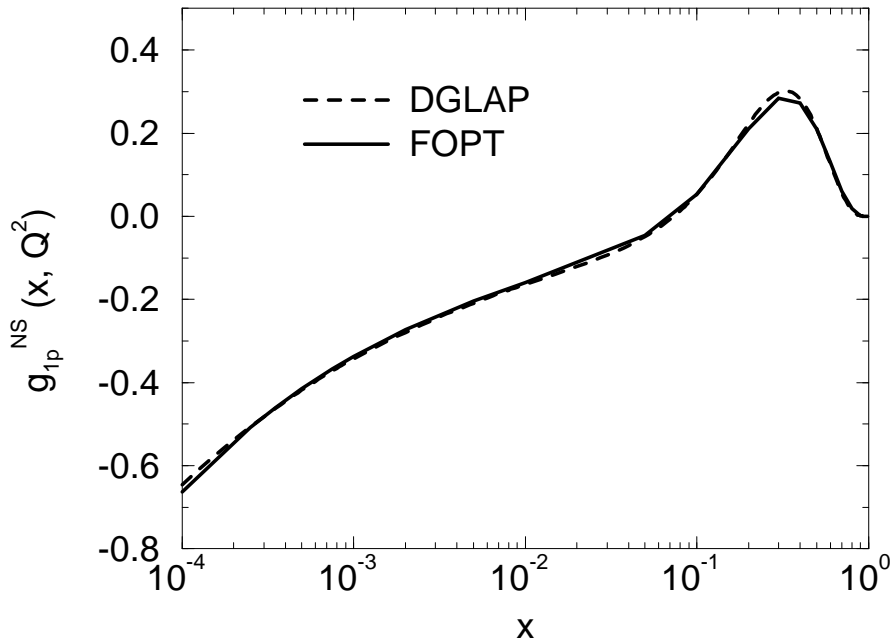}{20}{10}{-2}{-10}{0.9}
\caption{The non-singlet part of $g_{1p}$. The evolution is 
from 2.56 to 10 $GeV^2$. The continuous and dashed lines are,
respectively, the FOPT and DGLAP results at 10 $GeV^2$.}
\label{fig5}
\end{figure}

As in the unpolarized case, the polarized non-singlet part of 
$g_{1p} (x, Q^2)$ does not distinguish between FOPT and 
DGLAP evolution, as can be seen from Figure \ref{fig5}.
This feature persists when $Q^2$ is allowed to take values as
high as 1000 $GeV^2$, in a pattern very close to that of 
$F_{2p}^{NS}$. Hence,
in practical terms, no problem is detected with the
large logarithms involved: there is no practical 
necessity for their resummation.

\begin{figure}[htb]
\graphics{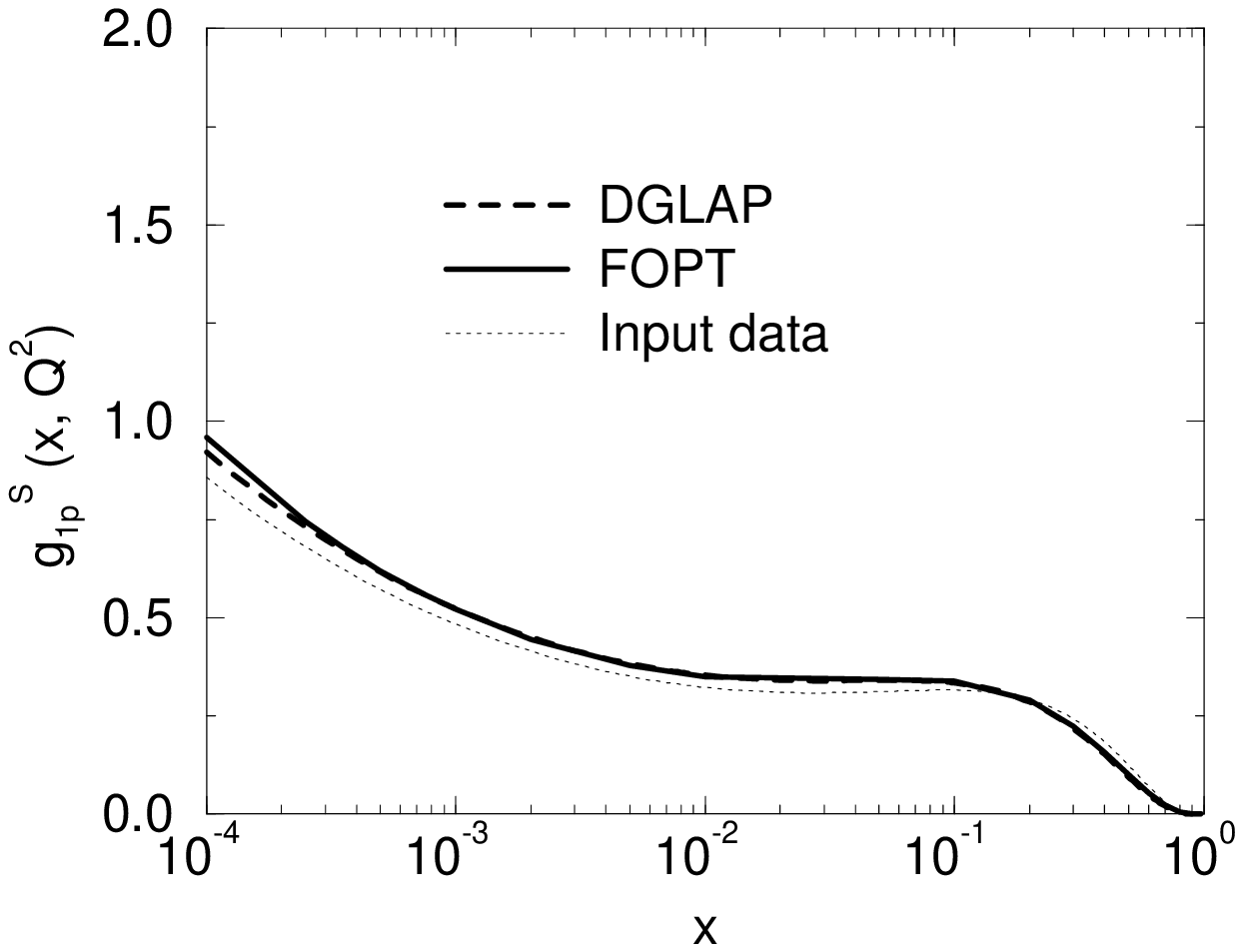}{20}{10}{-2}{-10}{0.9}
\caption{The singlet part of $g_{1p}$. The evolution is 
from 2.56 to 10 $GeV^2$. The continuous and dashed lines are,
respectively, the FOPT and DGLAP results at 10 $GeV^2$. The dotted
line is the input distribution at 2.56 $GeV^2$.}
\label{fig7}
\end{figure}

\begin{figure}[htb]
\graphics{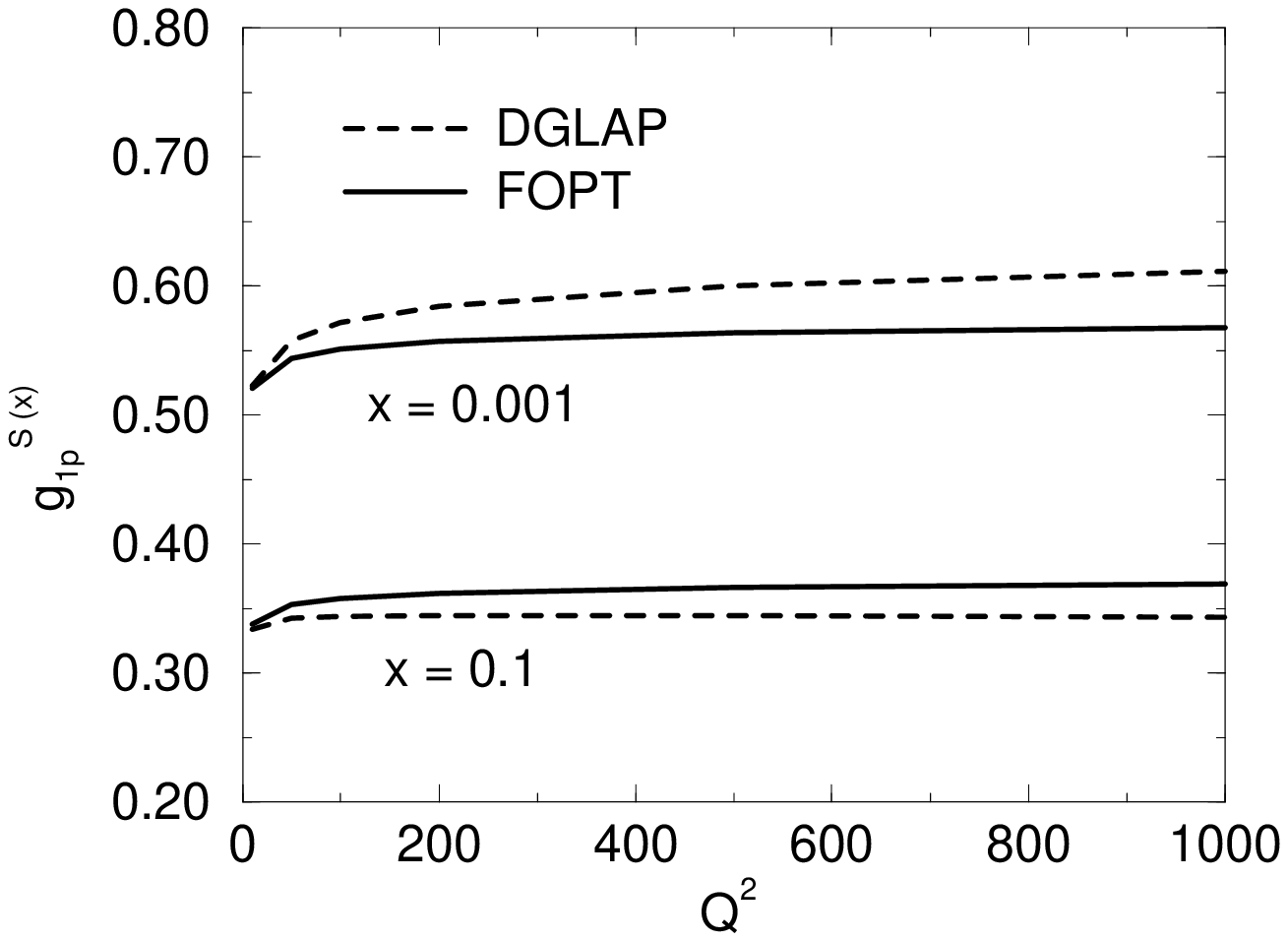}{20}{10}{-2}{-10}{0.9}
\caption{The $Q^2$ dependence of $g_{1p}^{S}$, for two specific
values of $x$. The initial scale for evolution is 2.56 $GeV^2$.}
\label{fig8}
\end{figure}

Surprisingly, in the singlet sector we still see the same
sort of behaviour.
Figure \ref{fig7} shows the $x$ dependence of $g_{1p}^S (x, Q^2)$
calculated at 10 $GeV^2$, and it is clear that the two methods of calculating
the evolution of $g_{1p} (x, Q^2)$ give identical answers. 
More astonishing, there is no substantial variation
with $Q^2$ even for small $x$ values as we can see in Figure \ref{fig8}, 
in a  behaviour completely distinct from the unpolarized case.

A possible interpretation of this result goes as follows.
In the unpolarized sector, there is a strong singularity in 
the gluon splitting function which governs the singlet DGLAP equations
for small enough $x$. The fact that FOPT can not match 
DGLAP in the evolution of $F_{2p}^S (x, Q^2)$ means that 
the singularities appearing in the Wilson coefficients Eqs. 
\ref{a4} and \ref{a5} are
not strong enough to match the singularities in  $P_{gg}$. Thus the
difference between FOPT and DGLAP singlet evolution appears, even
when the logarithms in the scale are small. On the other hand, the
nonsinglet sector has no strong small $x$ singularity in the splitting
functions, implying in the  agreement between FOPT and DGLAP in the 
whole $x$ region. Moreover, we observe that the difference 
between FOPT and DGLAP evolution grows mainly in a relatively
small $Q^2$ region, smaller than 100 $GeV^2$. Passing this region,
both evolutions become weakly dependent on $Q^2$. That would imply 
that the ``need'' for resummation of the $ln(Q^2/\mu^2)$ terms 
is more connected to the $x$ singularities in the splitting functions
than to the actual size of these scale logarithms.
  
For the polarized sector, the DGLAP evolution of the singlet   
distributions is governed not only by singularities in the 
gluon splitting function but also in the quark splitting function.
In fact, the polarized quark singlet and the polarized gluon
distributions evolve with different signs in the 
small $x$ region \cite{wally}.
The agreement between FOPT and DGLAP singlet evolution is, in this
sense, an indication that these singularities approximately cancel
when doing the DGLAP evolution.
This mechanism would allow  
FOPT evolution to follow DGLAP evolution closely, and it
also gives a fair description or justification
of why the unpolarized evolution is faster than its polarized counterpart:
there is a competition between the quark and gluon singularities, making
the final result for polarized structure function evolution slower.
Using a more physical language, the last statement should be synonymous 
with saying that not all of
the quark-antiquark pairs created by gluons in the small $x$ region
will be polarized\footnote{I thank Wally Melnitchouk for discussions on
this point.}. Hence unpolarized distributions would grow faster
than the polarized ones. In any case, a more detailed study on this
subject should be done. 

It may look contradictory that in an early work \cite{steffens} we
found that FOPT and DGLAP evolution gave quite 
different answers in the small $x$ region of the singlet piece of
$g_{1p} (x, Q^2)$. However, in that 
work we were considering heavy quark evolution using a zero
heavy quark distribution at $\mu^2$. Thus, the evolution was 
dominated by the gluon sector and the balance between the singularities in
the quark and in the gluon splitting functions, as discussed here, is no 
longer valid. As a simple check, we performed FOPT and DGLAP evolution
of $g_{1p}^S (x, Q^2)$ using the same gluon distribution used for
the evolution shown in 
Figure \ref{fig7} but using a zero for the quark singlet distribution.
As in Ref. \cite{steffens}, we find again a discrepancy between
the two methods, corroborating the scheme of canceling singularities.
We may then say that the reason we need to remove the mass 
singular terms from the heavy quark coefficient functions 
is in the absence of a heavy quark 
distribution at $\mu^2$, and not in the presence of mass singular 
logarithms. At least for polarized heavy quark electro production.

In summary, we have compared two methods of evolution for the structure
functions: FOPT and DGLAP evolution. We found that for the polarized structure
function $g_{1p} (x, Q^2)$, the two methods give identical answers, even
if the logarithms in the scale are large. For the unpolarized 
case, the two methods are not equivalent in the small $x$ region
of the singlet piece of $F_{2p} (x, Q^2)$. For the non-singlet piece, however,
there is no distinction between FOPT and DGLAP evolution, even for 
very large values of $Q^2$. The results of this work question the 
connection between large logarithms in $Q^2$ and the need for their 
resummation. 
\newline
\newline

I am very grateful to Wally Melnitchouk and to Fernando Navarra for 
many discussions on the subjects covered here. I would like to thank the 
Instituto de F\'{\i}sica
of the UFRGS (Brazil) for their kind hospitality, and where part of this 
work was completed. Thanks also goes to the Instituto de F\'{\i}sica 
Te\'orica (IFT - Brazil) for their continuous support and hospitality.
This work was supported by FAPESP (Brazil).

\addcontentsline{toc}{chapter}{\protect\numberline{}{References}}

\end{document}